\documentclass[letterpaper,twocolumn,10pt]{article}
\usepackage[dvips]{graphicx}
\usepackage{times}
\usepackage[small,compact]{titlesec}
\usepackage[small,it]{caption}
\usepackage{usenix}
\usepackage{endnotes}
\usepackage[tight]{subfigure}

\usepackage{url}
\usepackage{multirow}
\usepackage{array}
\usepackage{epsfig}
\usepackage{footnote}
\usepackage{amsmath}

\usepackage[compact]{titlesec}
\titlespacing{\section}{0pt}{*.5}{*.5}
\titlespacing{\subsection}{0pt}{*.5}{*.5}
\titlespacing{\subsubsection}{0pt}{*.5}{*.5}
\begin{document}

\title{\Large \bf On the Design and Implementation of Structured P2P VPNs}

\author{
David Isaac Wolinsky$^{\ast}$,
Linton Abraham$^{\bullet}$,
Kyungyong Lee$^{\ast}$,
Yonggang Liu$^{\ast}$,
\\
Jiangyan Xu$^{\ast}$,
P. Oscar Boykin$^{\ast}$,
Renato Figueiredo$^{\ast}$
\\
$^{\ast}$University of Florida, 
$^{\bullet}$Clemson University
\\
}


\twocolumn[%
\centerline{\Large \bf On the Design and Implementation of Structured P2P VPNs}

\medskip

\centerline{\bf 
  David Isaac Wolinsky$^{\ast}$,
  Linton Abraham$^{\bullet}$,
  Kyungyong Lee$^{\ast}$,
  Yonggang Liu$^{\ast}$,
}
\centerline{\bf
  Jiangyan Xu$^{\ast}$,
  P. Oscar Boykin$^{\ast}$,
  Renato Figueiredo$^{\ast}$
}
\centerline{
  $^{\ast}$University of Florida, 
  $^{\bullet}$Clemson University
}
\bigskip
]

\subsection*{Abstract}
Centralized Virtual Private Networks (VPNs) when used in distributed systems
have performance constraints as all traffic must traverse through a central server.
In recent years, there has been a paradigm shift towards the use of P2P in
VPNs to alleviate pressure placed upon the central server by allowing participants
to communicate directly with each other, relegating the server to handling session management and
supporting NAT traversal using relays when necessary.  Another, less common,
approach uses unstructured P2P systems to remove all centralization from the VPN.
These approaches currently lack the depth in security options provided
by other VPN solutions, and their scalability constraints have not been well studied.

In this paper, we propose and implement a novel VPN
architecture, which uses a structured P2P system for peer discovery, session
management, NAT traversal, and autonomic relay selection and a
central server as a partially-automated public key infrastructure (PKI)
via a user-friendly web interface.  Our model also provides the first design and
implementation of a P2P VPN with full tunneling support, whereby all non-P2P based
Internet traffic routes through a trusted third party and does so in a way that
is more secure than existing full tunnel techniques.  To verify our model,
we evaluate our reference implementation by comparing it quantitatively 
to other VPN technologies focusing on latency, bandwidth,
and memory usage.  We also discuss some of our experiences with developing,
maintaining, and deploying a P2P VPN.

\section{Introduction}
A Virtual Private Network (VPN) provides the illusion of a Local Area Network
(LAN) spanning a wide area network (WAN) infrastructure by creating
secure and authenticated communication links amongst participants.  Common uses of
VPNs include secure access to enterprise network resources from remote/insecure
locations, connecting distributed resources from multiple sites, and
establishing virtual LANs for multiplayer video games over the Internet.  In
the context of this paper, we focus on VPNs that provide connectivity amongst
individual resources each configured with VPN software.
Our work is significantly different
in scope from approaches that define VPNs ``as the `emulation of a private
Wide Area Network (WAN) facility using IP facilities' (including the public
Internet or private IP backbones).''~\cite{ip_vpns}.  The purpose of these
VPNs is to connect large sets of machines through virtual routers to a
virtual WAN environment.

The architecture described in this paper addresses a usage scenario where
participants desire VPN connectivity without incurring the complexity or
management cost of traditional VPNs. For instance, in a small/medium business
(SMB) environment, it is often desirable to interconnect desktops and servers
across distributed sites, provide authenticated access and encrypted
traffic to enterprise networked resources, and secure Internet traffic from mobile
users at untrusted locations.  Another example is collaborative academic environments
linking individuals belonging to a virtual organization
spanning multiple institutions, where coordinated configuration of network
infrastructure across different  sites is often impractical. Existing
approaches suffer from one or more limitations that hinder their
applicability in these scenarios: centralized approaches (e.g.
OpenVPN~\cite{openvpn}) incur the management cost of dedicated infrastructures.
Alternative P2P-based approaches (e.g. Hamachi~\cite{hamachi}) are
vulnerable to man-in-the-middle attacks if session management is handled by an
external provider; furthermore, they lack the support to tunnel traffic
to Internet hosts through the VPN.

Among existing VPN approaches
\cite{openvpn, hamachi, wippien, p2pvpn, n2n, tinc}, there is not a
single solution that provides the following in an integrated manner: 
no dependence on centralized server(s) for creation, maintenance and tear-down of VPN links;
completely secure full tunneling of Internet traffic;
decentralized relay selection, where any peer can potentially be a relay;
and user-friendly, intuitive membership management in a VPN.

Our system supports the well-known PKI-based security model to secure VPN links; 
for improved usability, we employ a semi-automated 
PKI managed through a group-based web interface.
The interface allows VPN group managers to review applications into the group and remove malicious
users.  Furthermore, we use the bootstrapping of a private P2P system off
an existing public P2P system to create a trusted P2P system that spans only the
members of a VPN group providing secure P2P routing and distributed hash table (DHT) storage.
We explore the problem of providing full tunneling to P2P VPNs and provide
a working solution that supports multiple gateways in a single VPN.  We
provide mechanisms that allow peers to establish NAT tunneling two-hop relay links based upon
node stability (uptime) and proximity (latency) when direct connectivity is unavailable (e.g. due to NAT traversal constraints).  The main
contribution of this paper is the design and evaluation of a novel structured P2P VPN overlay architecture.
While many of the individual components of our solution are not novel in
and of themselves, the integration of these components and their interaction results
in a unique system.

The rest of this paper is organized as follows.  Section~\ref{vpns} gives an overview
of current VPN technologies.
Section~\ref{structured_p2p} provides background on P2P systems.
Section~\ref{p2pvpn} describes the techniques used to create structured P2P VPNs.
In Section~\ref{evaluation}, we present evaluation
comparing our system with other VPNs.  Our experience developing, using, and debugging the
system is discussed in Section~\ref{experiences}. Section~\ref{related_work} discusses
related works, while Section~\ref{conclusions} presents concluding remarks.

\section{Virtual Private Networks}
\label{vpns}
There exist many different flavors of virtual networking.  This paper focuses on
those that are used to create or extend a virtual layer 3 network, of which there
are many types as summarized in Table~\ref{tab:vpn_types}.  A brief survey of popular
VPNs comprising different configurations is summarized in Table~\ref{tab:vpns}.
This section overviews core features of VPN designs, beginning with client configuration of VPNs, followed
by an analysis of different VPN server configurations as highlighted in
Table~\ref{tab:vpn_types}.

\begin{table}[ht]
\setlength{\itemsep}{0pt}
\setlength{\parskip}{0pt}
\centering
\begin{tabular}[c]{|m{2cm}|m{5cm}|} \hline
Type & Description \\ \hline
Centralized & Clients communicate through one or more servers which are statically
configured \\ \hline
Centralized Servers / P2P Clients & Servers provide authentication, session management, and
optionally relay traffic; peers may communicate directly with each
other via P2P links if NAT traversal succeeds\\ \hline
Decentralized Servers and Clients & No distinction between clients and servers;
each member in the system authenticates directly with each other; links between
members must be explicitly defined \\ \hline
Unstructured P2P & No distinction between clients and servers; members either know
the entire network or use broadcast to discover routes between each other \\ \hline
Structured P2P & No distinction between clients and servers; members are usually
within $O(\log N)$ hops of each other via a greedy routing algorithm; use 
distributed data store for discovery \\ \hline
\end{tabular}
\caption{VPN Classifications}
\label{tab:vpn_types}
\end{table}

\subsection{Client VPN Configuration}
\label{clientvpn}
\begin{figure}[ht]
\centering
\includegraphics[width=2.75in]{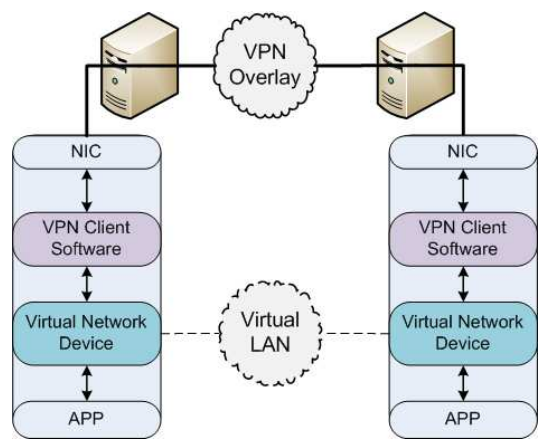}
\caption{A typical VPN client.  The VPN uses a VN device to make interaction
over the VPN transparent.  Packets going to VPN destinations are
directed towards the VN device, which interfaces  the VPN client.
The VPN client in turn sends and
receives packets over the hosts physical network device.}
\label{fig:vpn}
\end{figure}

In Figure~\ref{fig:vpn}, we abstract the common features of all VPNs clients.
The key components of a client machine are 1) client software that
communicates with the VPN overlay and 2) a virtual network (VN)
device.  During initialization, the VPN software starts by authenticating with an
overlay or VPN agent. Then, optionally, it queries the agent for information
about the network, such as the network address space, and finally the VN device
is started enabling secure communication amongst participants.

There are many different mechanisms for communicating with an overlay agent.
For quick setup, a system may require no authentication or use a shared secret
such as a key or password.  Using accounts and passwords with or without a shared
secret provides individualized authentication, allowing an administrator to block
all users if the shared secret is compromised or individual users if they act
maliciously.  For the strongest level of security, each client can be
configured to have a unique signed certificate that makes brute force attacks very
difficult.  The trade-offs come in terms of usability and management.  While the use of
signed certificates provides better security than shared secrets, it can be more
difficult to set up and use.  In a system comprising of non-experts, the
typical setup includes the use of
a shared secret and individual user accounts, where the shared secret is
included with the installation of the VPN application which is 
distributed from a secured site.

To communicate over the VPN transparently, a system must have a VN
device driver, which provides the mechanisms to inject incoming packets and retrieve outgoing
packets from the networking stack, enabling the use of common network APIs such as Berkeley Sockets allowing
existing application to work over the VPN without modification.
There are many
different types of VN devices, though due to our focus
on an open platform, we focus on TAP~\cite{tap}. TAP allows the creation of one
or more Virtual Ethernet and / or IP devices and is available for almost all
modern operating systems including Windows, Linux, Mac OS/X, BSD, and Solaris.
A TAP device presents itself as a character device providing read and write operations.
Incoming packets from the VN are written to the TAP device and the networking
stack in the OS delivers the packet to the appropriate socket.  Outgoing
packets from local sockets are read from the TAP device.

The VN device can be configured manually through static addressing or
dynamically through dynamic host configuration process (DHCP)~\cite{dhcp0}.  
Setting the IP address of the VN device causes the system to add a new
rule to the routing table that directs all packets sent to the
VPN address space to be sent to the VN device.  Packets are read from the
the TAP device, encrypted and sent to the overlay via the VPN client.
The overlay delivers the packet to another client or a server  with a 
VN stack enabled.  Received packets are decrypted, verified for authenticity,
and then written to the TAP device.  In most cases, the IP layer header remains
unchanged, while VPN configuration determines how the Ethernet header is handled.

\begin{figure}[ht]
\centering
\includegraphics[width=2.75in]{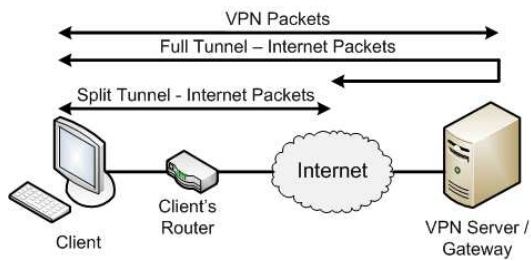}
\caption{A VPN setup expressing both full and split tunnel modes.  In both modes,
packets for the server are sent directly to the server.  In split
tunnel mode, Internet packets bypass the VPN and are routed directly to the Internet.  In full
tunnel mode, Internet packets are first routed to the VPN server / gateway, and then
to their Internet destination.}
\label{fig:tunnel}
\end{figure}

The described configuration so far creates what is known as a split tunnel: 
a VPN connection that handles \emph{internal VPN traffic only and
not Internet traffic}.  VPNs can also support
full tunneling, which allows a VPN client to securely forward \emph{all their Internet
traffic} through a VPN router.  This provides network-layer privacy and
authentication when a user is in an insecure environment, such as an open
wireless network at a coffee shop, by securely relaying all Internet traffic
through a trusted third party, the VPN gateway.  Both models are illustrated
in Figure~\ref{fig:tunnel}.

Most centralized VPNs implement full tunneling through a routing rule swap,
which makes the default gateway an endpoint in the VPN subnet and traffic
for the VPN server is routed explicitly to the LAN gateway.  For example, in a 
typical home network, an Internet bound packet will be retrieved at the VN
device, encrypted, and sent to the VPN gateway via the LAN's gateway.  At the
VPN gateway, the packet is decrypted and delivered to the Internet.  A P2P
system encounters two challenges in supporting full tunnels:  1) P2P traffic must not be routed
to the VPN gateway and 2) there may be more than one VPN gateway.  We further
discuss this issue and provide solutions to this problem in Section~\ref{fulltunnel}.

\begin{table*}[!h!t!]
\setlength{\itemsep}{0pt}
\setlength{\parskip}{0pt}
\centering
\begin{tabular}[c]{|m{2cm}||m{1.8cm}|m{2.4cm}|m{1.8cm}|m{2.9cm}|m{3.4cm}|} \hline
& VPN Type & Authentication Method & Peer Discovery & NAT Traversal & Availability \\ \hline \hline

OpenVPN~\cite{openvpn} & Centralized & Certificates or passwords with a central server &
Central server(s) & Relay through server(s) & Open Source\\ \hline

tinc~\cite{tinc} & Decentralized & PKI & Broadcast &
Relay through mesh & Open Source\\ \hline

CloudVPN~\cite{cloudvpn} & Decentralized & PKI & Broadcast &
Relay through mesh & Open Source\\ \hline

Hamachi~\cite{hamachi} & Centralized P2P & Password at central server & Central server &
NAT traversal and centralized relay & limited free-use, limited Non-Windows
clients, no private relays \\ \hline

GBridge~\cite{gbridge} & Centralized P2P & Password at central server & Central server &
NAT traversal, centralized relay & Windows only, freeware, no private relays \\ \hline

Wippien~\cite{wippien} & Centralized P2P & Password at central server & Central server & NAT traversal,
no relay support & Mixed Open / Closed source\\ \hline

N2N~\cite{n2n} & Unstructured P2P & Shared secret & Broadcast & NAT traversal,
decentralized relay & Open Source \\ \hline

P2PVPN~\cite{p2pvpn} & Unstructured P2P & Shared secret & Broadcast &
No NAT traversal, decentralized relay & Open Source \\ \hline

IPOP & Structured P2P & PKI or pre-exchanged keys &
DHT look up & NAT traversal and relay through physically close peers &
Open Source\\ \hline
\end{tabular}
\caption{VPN Comparison}
\label{tab:vpns}
\end{table*}

\subsection{Centralized VPN Servers}
OpenVPN is an open and well-documented platform for deploying
centralized VPNs. We use it as the basis for comparison with our approach,
as it provides a representation of features found in most centralized VPNs.
Centralized VPNs are responsible for authentication and routing between clients,
providing a NAT to the servers local resources and Internet (full tunnel), and
handling inter-server communication.

Central VPN servers operate at well-known endpoints 
consisting of a hostname or IP address and
a port.  In a system containing multiple servers, a client attempting to log in
will randomly attempt to connect to one of the servers until
successful, implementing a simple load balance.  Once connected, clients obtain an
address in the VPN address space.  Depending on configuration this will allow a
client to communicate with other clients, resources on the same network as the
server, or Internet hosts via the VPN.  In such situations, it is important that the
client and server both authenticate with each other in using some form of challenge
response protocol.

All inter-client communication flows through the central server.  In the default
configuration of OpenVPN, a client encrypts a packet and sends it to the server.
The server receives the packet, decrypts it, determines where to relay it, and then
encrypts and sends the packet to its destination.  This model does not
prevent a server from eavesdropping on such
communication.  While a second layer of encryption is possible through a shared
secret, it requires
out-of-band communication and is less secure than relying on a PKI.

To support full tunneling or allow the client to access the server's resources,
the server too must enable client-like features by becoming a VPN endpoint with
a VN device.  Depending on the configuration, the server can then configure
traffic from the VPN to go through a NAT prior to routing it to the LAN and/or
Internet.

OpenVPN allows a distribution of servers, so as to provide fault tolerance and
to a lesser degree load balancing.  Servers must be configured to know about
each other in advance and need routing rules established to forward packets.
Load balancing exists only in the process of the client randomly connecting to
different servers and potentially with a server refusing connection due to load.
The current approach lacks a distributed load balance.

Two examples of systems that assist in distributing load in VPN systems are
tinc~\cite{tinc} and CloudVPN~\cite{cloudvpn}.  Unlike the decentralized P2P
systems, these decentralized systems lack the ability to self-organize the VPN
and require explicit specification of which links to create.  This
means that, like OpenVPN, these systems can suffer VPN outages when nodes go offline.
The difference being that OpenVPN makes it explicit who is a server and who
is not, whereas in tinc and CloudVPN anyone can be a server or a client.  In the
typical tinc and CloudVPN setup, individual users share endpoints with each
other out of band and then place them in the VPN configuration file.  Due to
the lack of self-configuration, members in the system will not replace links
as members go offline.

\subsection{Centralized P2P VPN Systems}
Hamachi~\cite{hamachi} began the advent of centralized VPNs that went with the
ambiguous moniker ``P2P VPN''.  In reality, these systems would be best
classified as centralized VPN servers with P2P clients.  Specifically, the
nature of P2P in these~\cite{wippien, gbridge} types of systems provides direct
connectivity between clients once authenticated by a central server.  While
direct connection is desirable, it does not always happen due to firewalls or
impenetrable NATs. When this happens, the central server either acts as a relay,
if not, the two machines are unable to communicate.  One concern is
that each of these implementations uses their own security protocols that
involve using a server to verify the authenticity and setup secure connections
between clients.  Most of
these projects are closed preventing users from hosting their own
authentication servers and relays.  This model forces users to trust the third-party
server to not eavesdrop or perform other man-in-the-middle attacks.  None of these
provide support for full tunneling.

\subsection{P2P VPN Client / Server Roles}
\label{introp2p}
Unlike centralized systems, pure (or decentralized) P2P systems have no concept
of dedicated servers, though it is entirely possible to add reliability to the
system by starting dedicated instances of the P2P VPN.  In these systems, all
participants are members of a collective known as an overlay.  Current generation
P2P decentralized VPNs use unstructured P2P networks, where there are no
guarantees about distance and routability between peers.  Two popular examples of
unstructured P2P VPNs are N2N~\cite{n2n} and P2PVPN\footnote{Due to the similarities
between the name P2PVPN and focus of this paper, we use
``P2PVPN'' to refer only to ~\cite{p2pvpn} and ``P2P VPN'' to refer
explicitly to our approach.}~\cite{p2pvpn}.  As a result,
participants tend to be connected to a random distribution of peers in the
overlay.  Finding a peer requires some form of broadcasting, either announcements
or searches, to the entire overlay.  While
unstructured P2P systems have some scalability concerns, P2P systems in general
allow for server-less systems.  In the realm of VPNs, all client VPNs are also
servers with varying responsibilities depending on the VPN
application, as we present in Table~\ref{tab:vpns}.

Typically, decentralized, P2P VPNs begin by attempting to connect with well-known
endpoints running the P2P overlay software.  A list of such end points can be
maintained by occasionally querying the overlay for active participants
on public IP addresses and distributing the list with the application or some other
out-of-band mechanism.  In the case of P2PVPN, this involves communication with
one or more BitTorrent trackers to find other members of the P2PVPN group. 
N2N~\cite{n2n} requires knowledge of an existing peer in the system.  It uses
this endpoint to bootstrap more connections to other peers in the system,
allowing the application to be an active participant in the overlay and
 potentially be a bootstrap connection for other peers attempting to connect.

\section{Structured Peer-to-Peer Systems}
\label{structured_p2p}
Structured P2P systems provide distributed look up services with guaranteed
search time in $O(\log N)$ to $O(\log_2 N)$ time, in contrast to unstructured systems,
which rely on global knowledge/broadcasts, or stochastic techniques such as random walks
\cite{unstructured_v_structured}.  Some examples of structured systems can be found
in~\cite{pastry, chord, symphony, kademlia, can}.  In general, structured
systems are able to make these guarantees by self-organizing a structured topology such as a 2D ring or a hypercube.

The node ID, drawn from a large address space, must be unique to each peer, otherwise an address
collision occurs which can prevent nodes from participating in the overlay.  Furthermore,
having the node IDs well distributed assist in providing better scalability
as many algorithms for selection of shortcuts depend on having node IDs uniformly
distributed across the entire address space.  A simple mechanism to ensure this
is to have each node use a cryptographically strong random number
generator.    
Another mechanism for distributing node IDs involves the use
of a trusted third party to generate node IDs and cryptographically sign
them~\cite{secure_routing}.

As with unstructured P2P systems, in order for an incoming node to connect
with the system it must know of at least one active participant.
A list of nodes that are
running on public addresses should be maintained and distributed with the
application, available through some out-of-band mechanism, or possibly using
multicast to find pools~\cite{pastry}.  

Depending on the protocol, a node must be connected to either the closest
neighbor smaller, larger, or both.  Optimizations for fault tolerance suggest
that it should be between 2 to $\log(N)$ on both sides.  If a peer does not
know the address of its immediate predecessor or successor and a message
is routed through it destined for them, depending on the message type, it may
either be locally consumed or thrown away, never arriving at its appropriate
destination.  Thus having multiple
peers on both sides assist in stabilizing when the experiencing churn,
particularly when peers leave without warning.

Overlay shortcuts enable efficient routing in ring-structured P2P systems.  
The different shortcut selection methods include: maintaining large tables without
using connections and only verifying usability when routing
messages~\cite{pastry, kademlia}, maintaining a connection with a peer every
set distance in the P2P address space~\cite{chord}, or using locations drawn
from a harmonic distribution in the node address space~\cite{symphony}.

\section{Components of a P2P VPN}
\label{p2pvpn}
Before presenting our contributions, we first review our
current work as it provides the basis for our P2P VPN.  At the heart of our
system lies a P2P system similar to Symphony~\cite{symphony} named
Brunet~\cite{brunet}.  The specific components of the system that make it
interesting for use in a P2P VPN system include: STUN NAT traversal~\cite{stun},
system stability when two nodes next to each other in address space cannot
directly connect~\cite{hpdc08_0}, proximity-based selection of shortcuts~\cite{hpdc08_0},
a distributed data store based upon a DHT~\cite{pcgrid07},
and self-optimizing shortcuts to support single-hop connectivity between peers
when virtual IP traffic is detected between endpoints~\cite{wow}.

We have implemented a VN on top of the Brunet P2P library  with the
following features: self-configuring, low overhead use~\cite{sc09, ipop},
address configuration through a virtual DHCP server using DHT~\cite{sc09, pcgrid07},
ability to behave as a VN interface or router~\cite{sc09}, and
secure end-to-end (EtE) and point-to-point (PtP) links.
Our implementation is portable to any system that supports Tap and a C\# run-time (e.g. Mono, .Net),
and has been used in grid computing for over 3 years~\cite{archer,vtdc,pcgrid08,gridappliance}.

While this framework provides the basis for our design and implementation, we will show
in this paper that our approach generalizes to other structured P2P systems.

\subsection{Automating security configuration}
Our VPN supports the PKI model, where a centralized CA signs all client
certificates and clients can verify each other without CA interaction by using the CA's public
certificate.  However, setting up, deploying, and then maintaining security credentials can easily
become a non-negligible task, especially for non-experts.  Most PKI-enabled VPN systems 
require the use of command-line utilities, setting up your own methods of
securely deploying certificates and policing users. All these techniques can be applied in the P2P VPN of this paper, 
but our experience with real deployments of our system
indicates that usability is very important, which is why we sought a model with
easy to user interfaces.  In this section, we
present our solution, a partially automated PKI reliant on a redistributable group based web
interface.  Although this does not preclude other methods of CA interaction,
our experience has shown that it provides a model that is satisfactory for many use cases.

In order to obtain certificates, a user creates an account and joins a group through a Web interface,
providing relevant information as to the reason for joining the group. 
Users can also create VPN groups of their own through the interface, in effect becoming the administrator of the VPN group.
One or more group VPN administrators can verify this information and approve or deny the user
access into the group.  If a user is granted access, they are then able to
download a configuration blob.  The blob contains VPN network configuration
and a shared key that uniquely identifies the user and allowing them to
automatically retrieve signed certificates in the future.  The user configures the VPN
client with the blob.  Upon starting, the VPN client queries the group
server using the information in the blob to authenticate the server and itself to
the server.  After which, the user provides the group server a certificate
request containing the VPN client's node ID, which the server signs and returns.
At which point, the VPN client can connect to the VPN pool and communicate
securely with others in the group.  It is imperative that any operations that
involve the exchanging of secret information, such as the shared secret, be performed
over a secure transport, such as HTTPS, which can be done with no user
intervention.

Unlike decentralized systems that use shared secrets, in which the creator of
the VPN becomes powerless to control malicious users, a PKI enables the creator
to effectively remove malicious users.  The methods that we
have incorporated include:  use of a certificate revocation list (CRL) hosted
on the group server, DHT events requesting notification of peer removal from
the group, and broadcasting to the entire P2P system the revocation of the peer.

A CRL offers an out of band mechanism for distributing user revocations.
The CRL assist in cases where a malicious user can
use common P2P attacks to prevent notification of certificate revocations
transmitted via the overlay, as it is significantly more difficult to prevent
the retrieval of a CRL.

The DHT method acts as an event notification.  Upon a revocation,
the CA retrieves the values at a DHT key specific to the revoked peer.
The values contain node IDs of nodes who want to know of the revocation.
The problem with a
DHT is that it can be easily compromised if they have not been implemented
with significant measures to protect against maliciousness.

The most rudimentary mechanism is broadcasting the certificate
revocation over the entire P2P overlay.  In small networks, the cost of such a
broadcast may be negligible, but as a network grows, such a broadcast may
become prohibitively expensive.


\subsection{Bootstrapping Private Overlays}
In P2P systems, distributed security may not provide the same level of security
as centralized or managed security. A starting point to secure an overlay is to use
well-known security concepts such as PKI and SSL to encourage
wider adoption of P2P systems.  One problem with this approach though is that
users who want a private overlay may not have the resources, i.e. public
addresses, to host their own overlays.  To address this, we suggest
bootstrapping a private overlay from an existing public overlay~\cite{one_ring}.  Communication within the private
overlay is completely encrypted and authenticated with only members of the
VPN allowed access.

In this approach, members of the VPN are the only participants of the private overlay, providing a
model that is synergistic with the group VPN interface described earlier.  This
prevents malicious users outside of the VPN from attacking it, and facilitates 
the removal of misbehaving peers, primarily rooted in the fact that
the use of a broadcast to signal a certificate revocation is within the scope of the private overlay.

The process for bootstrapping a private overlay is as follows.  Once the VPN software
begins, it starts by connecting with the public overlay.  It queries the public
overlay's DHT at the key ``private:groupname'', where groupname is the
GroupVPN's name.  The values stored at the key are the public overlay node IDs
of the private overlays active members.  The joining VPN software will
attempt to form private overlay bootstrap connections with members of this
list using the public overlay.  During this process, both peers verify each
other's authenticity and form a secure connection.  This connection is then used
to bootstrap direct connections with members of the private overlay.  The reason
why the public overlay node IDs are stored at the public overlay's DHT key and
for using private overlay bootstrap connections over the public overlay is to
support NAT traversal.  This model supports reusing Brunet's underlying NAT
traversal techniques.  As a member of a
private overlay, a VPN node can elect to store information relevant to the VPN in the
private overlay's DHT, relegating the public overlay for private overlay
discovery.  VPN traffic can then be kept to the private overlay only or
the public overlay can be used for relaying, when
direct connectivity is not possible, though this has not been looked into yet.

\subsection{Full Tunneling over P2P}
\label{fulltunnel}
In full tunnel mode, all traffic to both VPN
and Internet hosts with the exception of VPN ``control'' messages route over
the VPN, as shown in Figure~\ref{fig:tunnel}.
In a centralized VPN, these ``control'' messages consist of
communication with the VPN server and full tunnel gateway.
In a P2P VPN, ``control'' messages consists of  communication between
Using full tunneling ensures that a malicious user cannot
easily eavesdrop into what would otherwise be public communication by
forwarding all non-VPN related traffic securely to a third party who resides
in a more trusted environment.  There are two key components to this scheme,
a gateway / server or traffic relayer and a client or a traffic forwarder.
In the following sections, we present methods for implementing
gateways and clients.

\subsubsection{The Gateway}
Configuring a machine as a gateway can be done with NAT software or using
IPOP's built-in NAT stack. For example, in Linux
this is possible through masquerading in iptables, which automatically handles forwarding
packets received on one interface to the next hop as well as receiving
packets from the Internet and forwarding them back to the appropriate client,
transparently taking care of NAT.

With a gateway intact, the VPN software can now announce that it provides
full tunneling.  For that purpose, we added an enable flag into the VPN
configuration to specify that a machine is a full tunnel gateway.  When the
VPN software starts, it will automatically append itself to the list of known
gateways for the VPN group in the DHT.

The only remaining difference in VPN gateway is the state machine used in
processing packets coming from the VPN.  Rather than rejecting packets
if the destination is not in the VPN subnet, they are only rejected if
gateway mode is disabled.  If it is enabled, all Internet packets are written to the
TAP device with the destination Ethernet address being that of the TAP device.
The remaining configuration is identical to other members of the system as
packets from the Internet to the client will automatically have the clients
IP as the destination as a product of the NAT.

\subsubsection{The Client}
\begin{figure}[ht]
\centering
\includegraphics[width=2.75in]{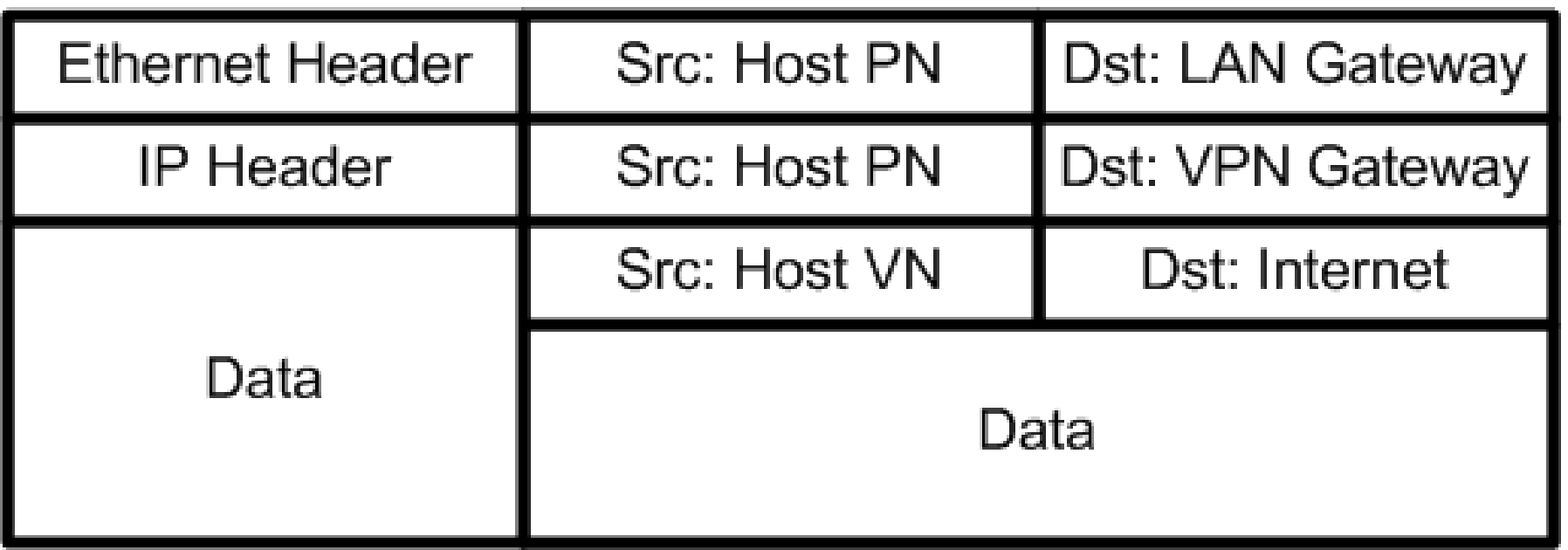}
\caption{The contents of a full tunnel Ethernet packet.  PN and VN are defined
as physical and virtual network, respectively.}
\label{fig:tunnel_packet}
\end{figure}

VPN Clients wishing to use full tunnel must redirect their default traffic to
their VN device.  In our VPN model, we use a virtual address for
the purpose of providing distributed VN services DHCP and DNS.  This same
address can be used as the VPN gateway, which works fine because as shown in
Figure~\ref{fig:tunnel_packet}, only the Ethernet header contains information
about the gateway.  Packets retrieved in the VPN software can then
forward the Internet packet to any machine in the VPN, providing gateway services
without changing IP or higher OSI layer changes.

The VPN's state machine has to be slightly modified to handle outgoing packets
not destined for a remote VPN end point.  Incoming packets destined for a subnet
outside of the VPN address space are rejected, unless full tunnel client mode
is enabled.   If it is enabled, the VPN software finds a remote peer to act as a
full tunnel gateway and then sends the packet to the remote peer.

The pathway for packets coming from the overlay needs to support full tunneling.
The source IP address will not match the VPN gateways IP but will be an Internet
address.  Thus the VPN client must confirm that the source is a VPN gateway, otherwise
the packet is thrown away.  Upon writing a packet to the VN device, the user
application will receive a packet from the Internet.

This leads to two issues: how to select the machine to use as a gateway and how
to configure a network stack to properly route packets and not direct all packets
to the VPN gateway.  Our model uses a simple mechanism for determining
which gateway to choose:  query the DHT for a list of potential gateways, select
a random one from the list, and verify liveness via periodic
ping messages.  For handling faults, we take a pessimistic approach, that is,
if a server is lost, we take note of it and only query the DHT again upon the
next outgoing Internet packet.

In this paper, we focus on the second issue: handling P2P-routed packets.
In the centralized VPN case, a client communicates directly
with a single point in the VPN, which is known ahead of time and can be implemented
by a simple routing rule swap prior to starting the VPN.
The same cannot be done for a P2P system.  Due to the dynamic, self-configuring nature of P2P systems,
ensuring that all P2P overlay messages are routed directly
becomes a non-trivial issue as there will be many such routes, most of which will
not be known ahead of time.  If we applied the same model used by centralized VPNs,
the client would end up routing both Internet and P2P traffic through the gateway
machine, creating a central point of failure.
We have considered two approaches, as outlined below.

\subsubsection{The Client -- Approach 1 -- Adding Routes}
The first approach is similar to the centralized version:
for each P2P link, we add an explicit rule in the
routing table so that packets destined for those endpoints are routed directly to them via
the LAN's default gateway.  In order to ensure this, we added a feature to the
socket handling code that would, prior to the first outgoing packet,
indirectly add a routing rule to direct packets for the remote node's
public address to the LAN gateway.  The rule would remain in effect until the
VPN closed down or the link was closed.
This model requires unique code for each OS platform; though supporting
Linux and Windows was quite trivial.
Because outgoing packets bear the source address of the physical network,
incoming packets will be delivered normally.

This method has two major shortcomings.  Common to all VPNs that employ
the standard route switch technique, all communication, not just VPN, is routed
directly to the server insecurely.  So while the VPN traffic is most likely
encrypted, a website hosted by the server might not be.
Another issue that arises in the P2P case is that
malicious users can send spoofed initiation packets
which will add extra routes to the routing table, resulting in a similar
situation as the first, unencrypted communication visible to eavesdroppers.
This led to a second approach described below.

\subsubsection{The Client -- Approach 2 -- Ethernet Frames}
This solution attempts to solve the problem introduced by the first solution,
removing any potential for eavesdropping.  In this solution all Internet packets with no
exception are directed to the VN device.  The VN is then responsible for filtering
P2P traffic, encapsulating them into Ethernet packets, and sending to the LAN's gateway.

In the VPN application, Incoming IP packets' source ports are compared to
VPN application's source ports.  Upon a match, the VPN application must
direct it to the LAN's gateway.  The three steps involved in this process are
1) translating the source IP address to match the physical Ethernet's IP address, 2)
encapsulating the IP packet in an Ethernet packet whose source is a random address
as described in~\cite{sc09} and destination is the LAN's gateway, and 3) sending the
packet via the physical Ethernet device.  The IP swap is required or the gateway will not
route the packet properly.  The issue of sending the Ethernet packet is not trivial as
Windows lacks support this operation.
Our platform independent solution uses a second TAP device bridged to the
physical Ethernet device, allowing Ethernet packets to be sent indirectly through the
Ethernet device via the TAP device.  This solution currently only works for UDP packets,
because all incoming packets will be directed towards the bridge and not the faked Ethernet
address.  TCP packets will send a TCP reset in this environment.
All other packets enter the system same as before.  This method has been verified
to work on both Linux and Windows using OS dependent TAP devices and bridge utilities.

While this method effectively resolves the lingering problem of ensuring that all packets
in a full tunnel will be secure, it raises a couple of issues 1) could the effect of
having all packets traverse the VPN application be prohibitively expensive and 2) why not
have the P2P application write LAN destined Ethernet packets directly avoiding the VPN client.
The cost of passing all packets through the VPN application is negligible, as shown in
Section~\ref{fulltunnel_eval},  since
they are destined for the wide area, where the time to traverse the network will be
orders of magnitude larger than passing through the VPN's packet processor.  Adding an
Ethernet writer to the P2P system creates an unclean abstraction as a VPN portion
of the system now becomes library dependent and reduces VPN software portability.

\subsection{Autonomic Relays}
When NAT traversal using STUN~\cite{stun} fails or there are other connectivity
issues some P2P VPNs~\cite{hamachi, gbridge} support relaying, similar
to Traversal Using Relay NAT (TURN)~\cite{turn} provided by a managed relay
infrastructure.
Centralized and decentralized VPNs do not suffer from this problem
as all traffic passes through the central server or managed links.  To address
the management and overhead concerns in these systems, we propose
the use of distributed, autonomic relaying system based upon previous
work~\cite{hpdc08_0,epost}.  Our previous work involved the use of
triangular routing that allowed peers next to each other in the
node ID space to communicate despite being unable to communicate directly
because of firewall, NAT, or Internet fragmentation issues.

\begin{figure}[ht]
\centering
\includegraphics[width=2.75in]{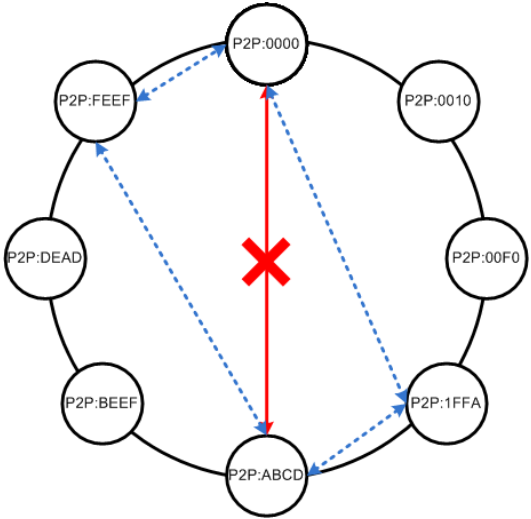}
\caption{Creating relays across the node address space, when direct
connectivity is not possible.  Two members, 0000 and ABCD,  desire a direct
connection but are unable to directly connect, perhaps due to NATs or firewalls.
They exchange neighbor information through the overlay and connect to one of
each other's neighbors, creating an overlap.  The overlap then becomes a 
relay path (represented by dashed lines), improving performance over routing
across the entire overlay.}
\label{fig:relay}
\end{figure}

The process for forming local relays or ''tunnels''~\cite{hpdc08_0} begins with
two nodes discovering
each other via existing peers and determining the need to be connected.
If a direct connection attempt fails, the peers exchange
neighbor sets through the overlay.  Upon receiving this list, the two peers
use the overlap in the neighbor sets to form a two-hop connection.  In this
work, we further extend this model to support cases when nodes do not
have an overlap set.  This involves having the peers connect to each other's
neighbor set proactively creating overlap, as represented in Figure~\ref{fig:relay}.

Additionally, we added the feature to exchange arbitrary information
along with the neighbor list.  So far we have implemented systems that pass
information about node stability (measured by the age of a connection) and proximity (based
upon ping latency to neighbors).  Additionally, when overlap changes, we make
it optional to select to use only a subset of the overlap, thus only the
fastest or most stable overlap is used with many more in reserve.

To verify the usefulness of two-hop over overlay routing,
we performed experiments and share the results in Section~\ref{relay_motivation}.
In a live system, we verify the accuracy and usefulness of our latency-based
relay selection algorithm in Section~\ref{relay_eval}.

\section{Evaluation of VPN Models}
\label{evaluation}
In this section, we evaluate our proposed P2P VPN
as described in Section~\ref{p2pvpn} implementing the features into
IPOP~\cite{sc09} and Brunet~\cite{brunet}.  We present the advantage of using
relays over overlay routing when NAT traversal does not work.  Then we examine
the effects of using different relay selection mechanisms.  Afterwards, we
evaluate the system overheads of OpenVPN, Hamachi, and our P2P VPN to determine
the OS resource costs and the cost of each in a distributed environment.
Finally, we present a quantitative comparison of our two full tunnel models.

\subsection{Motivation for Relays in the Overlay}
\label{relay_motivation}
The purpose of this experiment is to quantify the performance benefits of autonomic relays.
For this
experiment we use the MIT King data set~\cite{king_data}, a data set
containing all-to-all latencies between 1,740 well-distributed Internet hosts.
We reviewed many different sizes of networks up to 1,740 nodes, evaluating each
network size 100 times.  Our experiments were executed by running the VPN software in simulated mode,
which reuses the code base of IPOP to faithfully implement its functionality, but using 
event-driven simulated times to emulate WAN latencies in a LAN environment.  Once at steady state, we then calculate the average
all-to-all latency for all messages that would have taken two overlay hops
or more, the average of our low latency relay model, and the average of single
hop communication.  In the low latency relay model,
each destination node form a connection to the source node's physically
closest peer as determined via latency (in a live system by application level
ping).  Then this pathway is used as a two-hop relay between source and node.
We only look at two overlay hops and more, as a single hop would not necessarily
benefit from the work and would be the cause of a triangular inequality.  The
simulations were performed on a distributed grid platform, Archer~\cite{archer},
that uses IPOP for its virtual networking component. 

\begin{figure}[ht]
\centering
\includegraphics[width=3in]{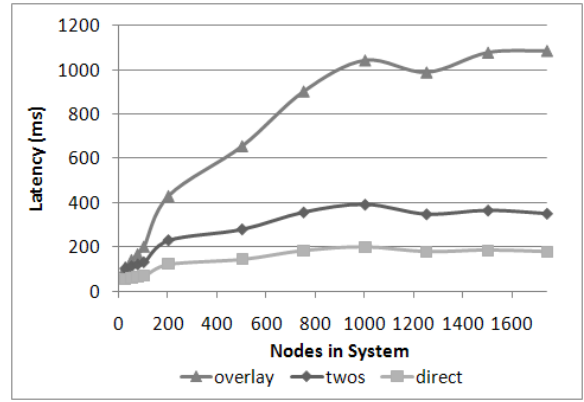}
\caption{A comparison of the average all-to-all overlay routing, two-hop relay, and
direct connection latency in a Structured P2P environment, Brunet, using the King
data set.}
\label{fig:simulated_relays}
\end{figure}

Our results are presented in Figure~\ref{fig:simulated_relays}.  We performed
the tests for varying network sizes.  We began our tests at 25, because network
sizes around 20 and under tend to be fully connected due to the connectivity
requirements of the system.  It is not until the network size expands past 100
and towards 200 nodes that relays become significantly beneficial.  At
100 nodes, there is approximately a 54\% performance increase, whereas at
200 there is an 87\% increase and it appears to grow proportionately to the size
of the pool.  The key take away is that latency-bound applications
using a reasonably sized overlay would significantly benefit from
the use of two-hop relays.  

\subsection{Comparing Relay Selection}
\label{relay_eval}
In this experiment, we share our experience of testing the use of latency-aware
relays using our public P2P pool running on Planet-Lab as well as Hamachi-Free
and Hamachi-Pro relays.  Due to Hamachi not supporting relays in Linux, this
experiment was performed in Windows Vista 64-bit.  The testing platform consists of
two virtual machine located on the same host with a firewall preventing them from
establishing direct connections.  All experiments were repeated 5 times using
a clean configuration each time.  In Hamachi, this meant that the server would
need to re-evaluate NAT traversing capabilities and the optimal relay to use.
In IPOP, this meant that the VPNs would generate a new node ID and become peers
with different members of the overlay.  Our results are presented in
Table~\ref{tab:relay_eval}.

\begin{table}[ht]
\setlength{\itemsep}{0pt}
\setlength{\parskip}{0pt}
\centering
\begin{tabular}[c]{|m{2.1cm}||m{.75cm}|m{.75cm}|m{.75cm}|m{.75cm}|} \hline
& \multicolumn{2}{|c|}{Latency} & \multicolumn{2}{|c|}{Bandwidth}\\ \hline \hline
& (ms) & stdev & Kbit/s & stdev \\ \hline
Hamachi-Free & 60.8 & 2.54 & 40.2 & 0.87 \\ \hline
Hamachi-Pro & 60.2 & 1.68 & 1000 & 1.29 \\ \hline
Latency-aware & 58.1 & 35.5 & 2245 & 1080 \\ \hline
\end{tabular}
\caption{Results of the evaluation comparing latency and bandwidth of Hamachi 
relays and IPOP latency-aware autonomic relay selection.}
\label{tab:relay_eval}
\end{table}

As Hamachi was started and figured out that NAT traversal was not possible, it
began using multiple different relays as evident by several different ping times.
Eventually Hamachi settled on a relay server and it appeared to be the same one
every time, for both Hamachi-Free and Hamachi-Pro.  The only difference between
Hamachi-Pro and Hamachi-Free is that in Pro there is a bandwidth cap of approximately
1 Mbit/s whereas Free is limited to 40 Kbit/s.  

The P2P system IPOP used has nodes both on Planet-Lab but also dedicated systems for
Archer~\cite{archer}.  These machines are at Universities and thus have a high bandwidth
and low latency connection to the testing site.  As witnessed by the results, it
appears that in most if not all these experiments peers had a low latency
connection to a University compute resource and it was chosen ahead of Planet-Lab.

The two apparent take aways are 1) the benefit of being able to deploy your own
relay servers and to reuse compute nodes as relay systems and 2)
as our network grows, we too may need to implement some form of bandwidth limit
at relay nodes.

\subsection{Comparing System Overheads}
In this experiment, we attempt to understand the bounds imposed by OpenVPN,
Hamachi, and IPOP.  We used Amazon EC2~\cite{ec2} to dynamically create various sized
networks ranging from 1 to 129 with one  client used as the control.
In the bootstrap phase, the control machine initiates communication with a
subset of the remaining clients in the VPN.  Once the system has warmed, 
the control continues pinging
this subset every 15 seconds for the next 10 minutes.  We capture bytes
transmitted into and out of the system, as well as the memory size of the
VPN application at the end of each stage.  As we test the varying network
sizes, we begin by communicating with 0 peers and exponentially increase the
amount (powers of two) until the control is communicating with all members of the VPN.
For these evaluations, we used a licensed version of Hamachi, but due to very recent
changes in Hamachi, the Linux client will not support networks larger than 16.
Due to amount of data collected and space limitations, we only present figures
for results that provide interesting data and summarize the rest in the text.

Memory, for the most part, exhibited an intuitive behavior: for each additional
connection there was more memory used.  The OpenVPN control client showed
negligible additional memory usage for additional nodes, though the server
showed a linear increment, around 1 MB, for each additional client in the system,
while activity had negligible effect on the results.  Hamachi had a
base line of a less than 1 MB and like OpenVPN, each additional client
in the system had a linear effect on memory, on the order of 4 KB, there was
no change based upon activity.  The effect of additional inactive nodes in
an IPOP network had negligible effect on memory, unlike Hamachi and OpenVPN.
The only time IPOP's memory consumption increased was during activity and it
scaled at a 200 KB per additional node.

\begin{figure}[ht]
\centering
\includegraphics[width=3in]{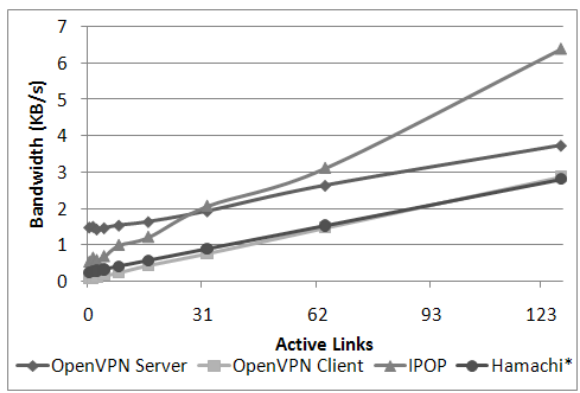}
\caption{Comparison of bandwidth costs for member activity versus network size.
As stated in the text, Hamachi limits the system to network sizes of 16, so to
estimate the Hamachi bandwidth we used the following formula using a linear
regression model based upon our data sets: $.049 + .002 KB/s * N + .02 KB/s * A$,
.049 the bandwidth in a network size of 0, i.e., keep-alives with the central
server, N is the network size, and A is the active links.}
\label{fig:bandwidth_128}
\end{figure}

In Figures~\ref{fig:bandwidth_128}, we present the bandwidth results for
client network sizes of 128.  Our evaluation scaled from having the control
communicate with 0 to 128 clients exponentially, though as mentioned already, Hamachi
only supports network sizes of 16, so we used a linear regression to extrapolate
the bandwidth results.  The results are somewhat predictable as Brunet, IPOP's
P2P infrastructure, primarily uses UDP connections and thus must maintain
application layer connections and NAT mappings.
It is not obvious that Hamachi is doing the same, which may be due to the fact
that the peers all have public IP addresses. In fact, Hamachi's behavior
appears to be similar to that of OpenVPN.  Additionally, OpenVPN does not need
to maintain NAT mappings as the servers are location on public addresses,
thus a client does not need to be as proactive about ensuring the connection
like in IPOP and Hamachi.

The important take-aways from this experiment are:  1) each additional member
in a VPN requires more memory, but the structured P2P approach places this burden
only on the individual client and only when there is active communication;
2) maintaining NAT mappings through keep-alive messages (a ping every 15
seconds) requires very little bandwidth; and 3) all traffic routes through the
OpenVPN server and it pays twice for each inter-client packet thus as
inter-client communication increases OpenVPN server's bandwidth becomes a
bottleneck.

\subsection{Full Tunnel Overhead}
\label{fulltunnel_eval}
This experiment was done to evaluate the overhead introduced using our Ethernet
packet based full tunnel technique.  The environment for this experiment consists
of a VPN client and gateway located at a residence and  University, respectively.
The gateway is configured using Linux with iptables' masquerading.  The client
is also using Linux.  The test performed consist of calculating the average
latency, through a minutes worth of ping measurements, between the client and
Google as well as the gateway's public and private addresses.

\begin{table}[ht]
\setlength{\itemsep}{0pt}
\setlength{\parskip}{0pt}
\centering
\begin{tabular}[c]{|m{1.5cm}||m{1.25cm}|m{1.25cm}|m{1.25cm}|} \hline
& Google & GW Pri & GW Pub \\ \hline \hline
Ethernet & 70.6 & 12.9 & 13.9 \\ \hline
Routing & 71.4 & 13.2 & 11.0 \\ \hline
None & 66.1 & N/A & 10.9 \\ \hline
\end{tabular}
\caption{Latency results comparing full tunnel approaches measured in ms.
GW Pri is the gateways VPN address, GW Pub is the gateways public
address, Ethernet was the Ethernet packet writing full tunnel method,
routing was the routing table manipulation method, and none is the baseline
without any VPN software.}
\label{tab:fulltunnel_eval}
\end{table}

As shown in Table~\ref{tab:fulltunnel_eval}, there is not a significant
latency difference in the full tunnel approaches.  One interesting result
is the latency to gateways public address in the routing test, which is a
result of the ping being sent insecurely avoiding the VPN stack completely.

\section{Experiences}
\label{experiences}
Our work on the P2P VPN model presented in this paper is based on four
years of work spent developing, deploying, and maintaining VN software for
grid computing~\cite{wow,sc09,archer}.  In this section, we first present
our different deployments of our P2P VPN and then our experiences
in debugging such a structured P2P VPN system.

\subsection{Deployments}
Initially IPOP was designed to assist in self-configuring ad hoc grid deployments
using virtualization, as part of the ``Grid Appliance'' project
\cite{gridappliance}.  Over the years, the Grid Appliance has matured and so
has our usage of IPOP.  Currently, the Grid Appliance is used as the basis
for a free-to-join computer architecture consisting of over 400 compute resources
spanning 5 seed universities with users across the United States and abroad.
The Grid Appliance allows easy creation of ad hoc, distributed grids through
the use of the IPOP and the structured P2P DHT.  Experienced users can use the
system to create large grids in less than an hour with most of the time spent
waiting for virtual machines to boot.  Non-expert users can quickly access the
distributed grid in a matter of minutes by downloading a virtual machine
manager and the Grid Appliance.  The comment we receive most often in our
polls is ``I was surprised at how it just works.''

Motivation for a significant portion of our work lies in a desire to integrate
other resources, not only virtual appliances,
in the same ``just works'' manner. The GroupVPN described in this paper is the same
software stack that runs inside Grid Appliance but also allows services the Grid
Appliance to connect with external NFS and external grid resources.

Our experience with P2P VPN software also includes SocialVPN~\cite{cops08}.  The
SocialVPN places each client in their own private network, where they are only
addressable by their friends as defined by third party social network.
Unlike GroupVPN, this makes each user the master
of who does and does not have access to their resources.  While SocialVPN
targets the creation of ``personal'' VPNs, which are often asymmetrical,
it does not lend itself well to environments that require some form of
central management and symmetric connectivity such as a computing grid or a
small to medium businesses.

\subsection{Discovering Faults}
In this section, we share our
experiences in managing and finding faults in a deployed P2P VPN overlay bootstrap 
service we run on Planet-Lab~\cite{planetlab}. Planet-Lab serves as a
challenging environment that stresses our system and helps us assess its reliability, stability, and
consistency.  The basic consistency check we use to discover faults is to compare peers knowledge of the pool and test
each peer for congruence with both its first and second left and right neighbors.
We call this a crawl.  This information is stored in a database, which we can
later use to find nodes that have been inconsistent many consecutive times.
If a node is able to fix inconsistencies in future crawls, experience suggests
the inconsistency was probably due to churn in the system.  Otherwise, it will
probably still be in an inconsistent state and we are able to query the node
and other nodes nearby for additional state information.  At a minimum, this
would be help us determine if there exists a problem and potentially the stale
state causing the connectivity issue.  If this reveals little
information, we either retrieve a log or request it from the user who owns node.
The log typically provides some useful information.  Additionally, as with all
multithreaded applications, deadlocks happen, we found it useful to add
liveness states to threads to assist in finding deadlocks.

Other information we monitor includes peer count, memory, and CPU usage.
Node count can be quite difficult to keep track of in Planet-Lab as machines at
a rate of 5 to 20 per day are restarted and our software is not automatically
restarted on these machines, thus the case to watch for is non-linear loss of
nodes.  Planet-Lab also places challenges on memory, as the systems can often be
I/O starved causing what appears to be memory leaks as Brunet's internal queue
can grow without bound.  In these cases, we have the node disconnect from the
overlay and sleep before returning.  The advantage of Planet-Lab as a test ground
is that it presents so many unique situations that can be very difficult to
reproduce in a lab controlled test system.  It is our belief
that any system that uses large scale Planet-Lab deployments as a testing ground
will be quite reliable.

We are still actively seeking better ways to check and verify the state of our system.
For example, the cost of doing a crawl can take $O(N \log(N))$ time,
since we have to communicate with every single node with an average routing
time of $O(\log(N))$.  Future work in the arena is focused on Brunet's
MapReduce~\cite{mapreduce} framework, which can be used to
provide system wide searches and status checks in $O(\log(N))$ time.

\section{Related Works}
\label{related_work}
Our work is not the first to propose using a group like mechanism for regulating
members in a VPN. Hamachi presently offers the ability to create and join a
network from either a VPN client or through their Web site.  To create a group,
a user can either form a private invite only VPN or a password protected public VPN.
Users must exchange out of band both the password and the VPNs name and both
registered and unregistered users (guest) can join Hamachi VPNs.  Also, Hamachi
uses a model similar to a key distribution center (KDC) as opposed to a PKI,
thus it is quite
easy for Hamachi to do man-in-the-middle attacks.  Hamachi's server is not
redistributable and all users must use LogMeIn's (Hamachi's owner) KDC and relays.
Our model ensures that each user is traceable and has to be authenticated by the
group administrator.  The groups is further secured by giving each user a
unique key to retrieve signed certificates.  Most importantly our PKI is open
and redistributable, so users can self-host the group VPN web server, and our
relays are built into the VPN and require no management.

Our group system is not the first to provide an automatic PKI. Previous work in
this field includes RobotCA~\cite{robotca}.  A RobotCAs receives request via e-mail,
verifies that the sender's e-mail address and embedded PGP key match, signs the
request, and mails it back to the sender.
RobotCAs are only as secure as the underlying e-mail infrastructure and provide
no guarantees about the person beyond their ownership of an e-mail address.  In
certain cases, this model could be used in our use cases, such as a SMB or for
universities if it enforces that all users use university e-mail addresses,
then the RobotCA 
would only sign e-mails if they come from a specific domain.  Our experience
suggests that is not rare for an academic to use a non-university e-mail for
university purposes.  Another concern is that the RobotCA would require management
to limit allowed users to members of a class or an organization whose e-mail
addresses does not contain domain names.

VINI~\cite{vini}, a network infrastructure for evaluating new protocols and
services, uses OpenVPN along with Click~\cite{click} to provide access
from a VINI instance to outside hosts, as an ingress mechanism.  OpenVPN only
supports a single server and gateway per a client
and does support distributed load balancing.  VINI may benefit from using a
VPN that uses a full tunnel model similar to ours, as it lends itself
readily to interesting load balancing schemes.

Our work is not the first to suggest using a P2P infrastructure to enable the
discovery of physically close TURN-like relays.  Skype~\cite{skype_overview}
queries super nodes in an attempt to find physically close relays.
The primary difference in our work is that our model could easily be configured
to let users create and select their own relays.

\subsection{P2P VPN in Other Structured Overlays}
\begin{table*}[!h!t]
\setlength{\itemsep}{0pt}
\setlength{\parskip}{0pt}
\centering
\begin{tabular}[c]{|m{2cm}||m{2.75cm}|m{2.8cm}|m{2.95cm}|m{1.65cm}|m{1.65cm}|} \hline
System & Overlay messaging & NAT Traversal & DHT & Secure PtP & Secure EtE\\ \hline \hline
Brunet & Yes (AHSender) & UDP and overlay relaying & Yes with reliability & Yes & Yes \\ \hline
FreePastry & Yes (route) & Only with port-forwarding enabled  & Yes, PAST~\cite{past}, reliable & No & No \\ \hline
NChord & No, only look up & No & Simple, non-fault tolerant DHT & No & No \\ \hline
\end{tabular}
\caption{A comparison of structured P2P systems.  PtP stands for point-to-point
communication, such as communication between physical connections in a P2P
overlay.  EtE stands for end-to-end communication, such as messages routed
over the overlay between two peers.}
\label{tab:structured_p2p_compare}
\end{table*}

The purpose of this work is to develop a P2P VPN model that can easily be
applied to other structured P2P systems.  In this section, we focus on the
portability of our platform to other structured P2P systems, namely
Pastry and Chord by analyzing FreePastry and
NChord respectively.  FreePastry can easily reuse our C\# implemented
library through the use of IKVM.NET, which allows the porting of
Java code into the CLR.  NChord is a Chord implementation written in C\#.  In
Table~\ref{tab:structured_p2p_compare}, we compare the features of the structured
P2P systems as they apply to the use as a VPN.
The specific focus of this section is to understand how discovery and VPN connections
would work.  Our discovery model for mapping node ID to IP works through the use of a DHT.
During the IP address allocation, the VPN client will place in the
DHT a key, value pair mapping a virtual IP to a node ID, as described in~\cite{sc09}.

The bootstrapping of a connection in NChord begins by finding the owner of the
DHT key containing the mapping of virtual IP address to node ID
through ``find\_successor''.  After establishing a connection with the owner of
that key, the owner needs to query for its value, which would be the node ID of the
node owning the virtual IP.  After executing ``find\_successor'' and retrieving
the destination nodes physical IP address, the VPN
can ``connect'' to the remote node using either a UDP or TCP socket.  Virtual IP
messages would then be sent and received through this ``connection''.  Unlike
our system, NChord does not support sending messages through the overlay,
thus a separate ``connection'' for application purposes will need to be
created.

FreePastry begins by looking up the mapping of virtual
IP to node ID in PAST.  To form a connection through overlay, the VPN
client would use ``route'' to send packets and ``deliver'' to receive incoming packets.
In fact, the model is very similar to Brunet.  Furthermore,  FreePastry has
knowledge of proximity in shortcuts, so it may be very easy to apply
high-performance autonomic relays in pastry potentially reusing some of their work from ePost~\cite{epost}.
FreePastry does not form shortcuts based upon communication demands, so if two peers were
actively communicating, they may always have to route traffic over the
overlay, which in most cases will be significantly slower than if they were
directly connected.  Though one could argue that since FreePastry does
not support NAT traversal, all nodes will already be public and thus an
application, as in the NChord example, could form a direct connection bypassing
the overlay.

\section{Conclusions}
\label{conclusions}
This paper presents a novel VPN approach that has been designed from the ground up
to facilitate ease of use while maintaining a reasonable level of security.  At
the core of the VPN is a structured P2P system that provides decentralized
peer discovery, NAT traversal, relays, full tunnel clients and gateways, and secure point-to-point
(PtP) and end-to-end (EtE) links.  In the paper, we focus primarily on relays,
full tunnels, and group security.  The relay seekers can use arbitrary information,
such as connection age or latency, to assist in finding suitable overlap.  With
overlap sets formed, relays will be created automatically using an existing infrastructure.
We present a novel method for client-side full tunneling that works in P2P and provides a level
of security not present in prior works by ensuring that only secure data is transmitted over the LAN.
The PtP and EtE links are secured via a group system
using a partially automated PKI with an intuitive web interface, which can be hosted
by a group or they may use our public system~\cite{gridappliance}.

Through the use of the group interface, users can create their own VPNs with
private P2P pools while using public pools as a bootstrap.  Users are able to join
the system by downloading a stock VPN and loading it with their user specific configuration or binary blob.  The
VPN will automatically connect to the appropriate web server and both parties will
verify each other's authenticity.  The user is then quickly connected to the VPN.
Administrators are able to police the system through the web interface and
offending users are eventually removed from the system through one of the following
mechanisms: CRL, DHT events, and broadcast messages.

The paper introduces many new interesting research problems: 1) distributed load 
balancing of full tunnel gateways, 2) understanding the long term effects of different
automatic relay selection models, 3) understanding the benefits beyond security
of using a private overlay for a VPN.  We believe that this work provides both a
position and a model on how to design VPNs.

\bibliographystyle{abbrv}
\small{
\bibliography{nsdi10}
\suppressfloats
}

\end{document}